\documentclass[fleqn,12pt,twoside]{article}
\usepackage{espcrc1,epsfig}

\usepackage{graphicx}
\usepackage[figuresright]{rotating}

\newcommand{\AmS}{{\protect\the\textfont2
  A\kern-.1667em\lower.5ex\hbox{M}\kern-.125emS}}

\title{High $p_T$ identified hadron ratios in $\sqrt{s_{NN}}$=200\,GeV Au+Au Collisions}

\author{T. Sakaguchi\address[CNS]{Center for Nuclear Study, Graduate School of Science, University of Tokyo, Wako Br. RIKEN, 2-1 Hirosawa, Wako, Saitama 351-0198, Japan} for the PHENIX Collaboration\footnote{for the full PHENIX Collaboration author list and acknowledgements, see Appendix "Collaborations" of this volume.}}
 
\begin{document}

% typeset front matter
\maketitle

\begin{abstract}
The PHENIX detector at RHIC measured high $p_T$ identified hadron ratios in
$\sqrt{s_{NN}}$ = 200\,GeV Au+Au collisions. Within the current systematic and
statistical errors, $\overline{p}/p$ ratios that are measured up to 3.8\,GeV/c
are almost independent of both $p_T$ and centrality. The baryon to meson ratio
is measured through $p/\pi$ and $\overline{p}/\pi$ ratios up to 3.8\,GeV/c,
showing they are strongly centrality dependent.
\end{abstract}

\section{Introduction}

The PHENIX measurement of high-$p_T$ suppression of neutral pions and
inclusive charged hadrons reveals flavor dependence in the strength of
the suppression~\cite{ppg003ppg006}.
Identified charged particle spectra hint that the difference is due to the 
unexpectedly high yields of high-$p_{T}$ ($>2GeV/c$) (anti)protons that appear
to be unsuppressed. Studying the particle composition at high-$p_{T}$ is
an important step in understanding baryon production and transport, system
evolution, and the interplay between soft and hard processes.

The high statistics $\sqrt{s_{NN}}$ = 200\,GeV Au+Au data set obtained by PHENIX
in RHIC RUN-2 and the results from $\sqrt{s_{NN}}$ = 130\,GeV Au+Au data in
RUN-1~\cite{130GeV} realized the systematic measurement of the ratio of baryon
to meson through the ratios of $p$($\overline{p}$) to $\pi$. The $\overline{p}$ to $p$ ratios
were also measured up to 3.8\,GeV/$c$ as a function of centrality. This paper
reports the results on the high $p_T$ identified hadron ratios in $\sqrt{s_{NN}}$=200\,GeV Au+Au 
collisions.

\section{Experimental Setup and Data analysis}
The PHENIX detector~\cite{phenixnim} has the capability of identifying charged
hadrons and neutral pions over a broad momentum range. The charged hadrons are tracked by a drift 
chamber followed by a pad chamber. Time-of-flight measurement using Beam-Beam counters (BBC) and a 
TOF-wall provides $\pi^{+/-}$ identification up to $p_{T}$ = 2\,GeV/$c$, and $p$($\overline{p}$)   
up to $p_{T}$ = 4\,GeV/$c$. Neutral pions are measured by an electromagnetic calorimeter via
$\pi^0 \rightarrow \gamma \gamma$ decay channel over the $p_T$ range of 1\,GeV/$c$$\le p_{T} \le$10\,GeV/$c$. Combining the two techniques, $p$($\overline{p}$)$/\pi$ and $\overline{p}/p$ ratios are obtained in the range of
0.5\,GeV/$c$$\le p_{T} \le$3.8\,GeV/$c$. The acceptance of the PHENIX detectors
used in this analysis are $\pm$0.35 in pseudo-rapidity and $\pi/4$ for
charged hadron identification and $\pi$ for neutral pions in azimuth,
respectively.

Minimum bias events are triggered and classified into centrality selections
using BBC and Zero-Degree calorimeters. The trigger efficiency is evaluated
by simulation to be 91.4\,\% of the inclusive inelastic cross-section.
The analyzed data set includes
 around 4\,M Au+Au minimum bias events for the charged hadron analysis and
30\,M for the neutral pion analysis.

Charged particle spectra were corrected for acceptance, decay in flight 
and tracking efficiency using a single particle Monte Carlo. Multiplicity-dependent 
corrections were evaluated by embedding simulated tracks into real events.  
The $p_T$-dependent part of the systematic errors on the measurement of
$\pi^+$($\pi^-$) and $p$($\overline{p}$) are 7\,\% and 11\,\%,
respectively~\cite{ref5}. These include uncertainties in particle identification and detector 
fiducial area cuts. The $p_T$-independent part of the errors that are dominated by the uncertainty in the multiplicity-dependent corrections are 14\,\% and 18\,\%
for $\pi^+$($\pi^-$) and $p$($\overline{p}$) for central events, while 14\,\%
and 16.4\,\% for peripheral events, respectively. By taking the ratios of
anti-particle to particle, the $p_T$-independent errors cancel out.
The $p_T$-dependent part of the systematic errors on $\pi^0$ measurement is
20-30\,\%, and the $p_T$-independent part is 9\,\%. Details on the contributing sources are given 
elsewhere~\cite{ref6}.

\section{Results}
\subsection{$\overline{p}/p$ ratios}
\begin{figure}[htbp]
\begin{center}
\vspace{-5mm}
\epsfig{file=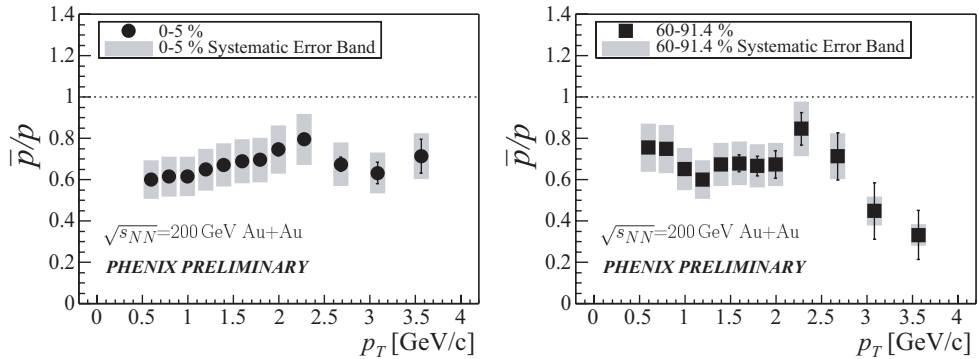,width=13cm}
\vspace{-10mm}
\caption{$\overline{p}/p$ ratios as a function of $p_T$. Left panel shows the ratio for 0-5\,\% centrality, and right panel shows that for 60-91.4\,\% centrality. The error bars indicate the statistical errors, and gray bands indicate the systematic errors for each data points. The ratio is almost flat over entire $p_T$ for central events, while it decreases at high $p_T$ for peripheral events. \label{fig:PbarP_ratio}}
\vspace{-8mm}
\end{center}
\end{figure}

\begin{figure}[htbp]
\begin{center}
\epsfig{file=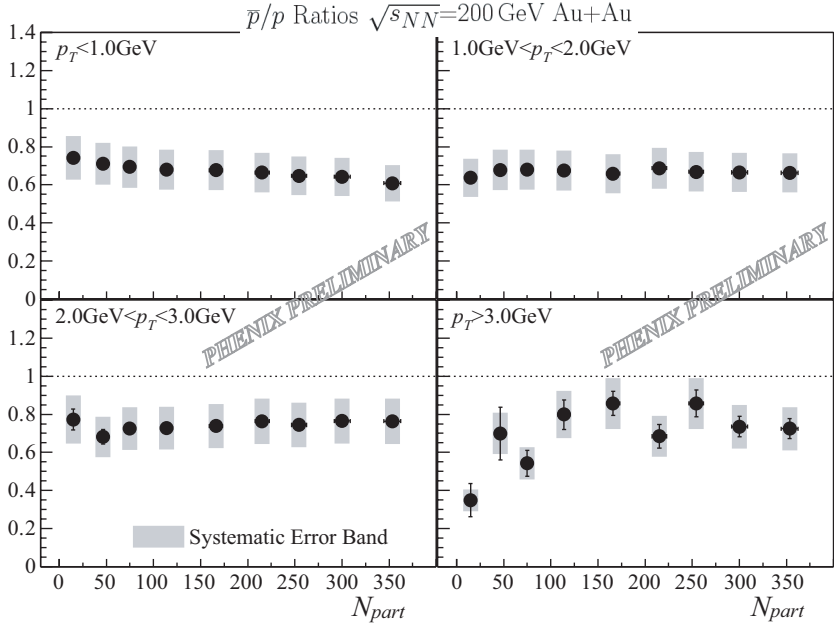,width=11cm}
\vspace{-10mm}
\caption{$\overline{p}/p$ ratios as a function of number of participant nucleons for different $p_T$ ranges. From top left to right bottom panels show the ratio for $p_T$$<$1.0\,GeV, 1.0\,GeV$<$$p_T$$<$2.0\,GeV, 2.0\,GeV$<$$p_T$$<$3.0\,GeV and $p_T$$>$3.0\,GeV, respectively. The definition of error bars is same as Fig.~\ref{fig:PbarP_ratio}. The ratios are almost flat over centralities. \label{fig:PbarP_Ncoll}}
\vspace{-14mm}
\end{center}
\end{figure}

Figure~\ref{fig:PbarP_ratio} shows the $\overline{p}/p$ ratios as a function
of $p_T$ for 0-5\,\% and 60-91.4\,\% centrality selections. The error bars show
the statistical errors, and the gray error bands show the systematic errors on
the data points. The ratio for central events is almost flat over
0.5\,GeV/$c$$<$$p_T$$<$3.8\,GeV/$c$ within the current systematic and
statistical errors, while it decreases at high $p_T$ for peripheral events.
For a detailed study, the ratios are also evaluated as a function of centrality
for four different $p_T$ ranges. Figure~\ref{fig:PbarP_Ncoll} shows the
$\overline{p}/p$ ratios as a function of number of participants which are
associated with centrality using a Glauber model calculation.
The error bars show the statistical errors, and the gray error bands show
the systematic errors. It can be concluded that the $\overline{p}/p$ ratios
are almost independent of centrality within the current systematic and
statistical errors.

\subsection{$\overline{p}/\pi$ and $p/\pi$ ratios}
\begin{figure}[htbp]
\begin{center}
\vspace{-6mm}
\epsfig{file=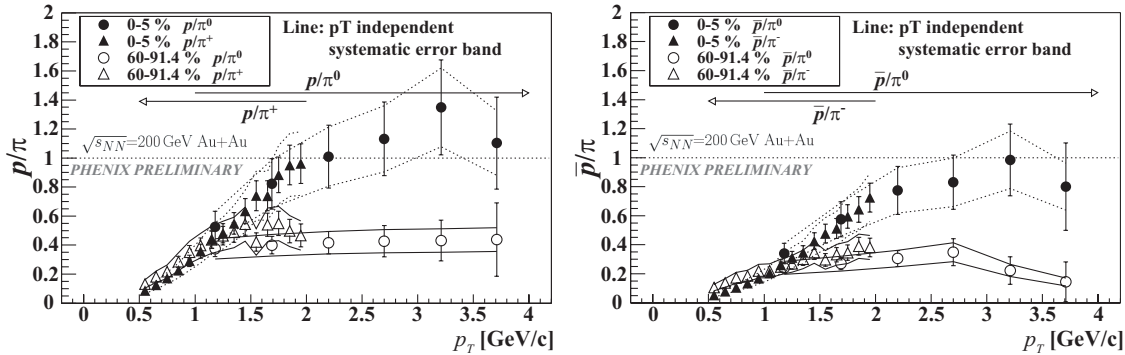,width=15cm}
\vspace{-10mm}
\caption{$p/\pi$ (left panel) and $\overline{p}/\pi$ (right panel) ratios as a function of $p_T$ for 0-5\,\% and 60-91.4\,\% centralities, respectively. Lines show the $p_T$-independent systematic error bands, and the error bars show the quadrature sum of statistical and $p_T$-dependent systematic errors. Both ratios in central collisions reach to unity at 2$\sim$3\,GeV/$c$. \label{fig:P_Pi}}
\vspace{-10mm}
\end{center}
\end{figure}

Figure~\ref{fig:P_Pi} shows the $p/\pi$ and $\overline{p}/\pi$ ratios as a
function of $p_T$ for two different centralities.
For the $p/\pi$ ratio, $p/\pi^+$ is plotted below 2\,GeV/$c$, and $p/\pi^0$
is overlayed above 1\,GeV/$c$ to obtain the ratio from 0.5 to 3.8\,GeV/$c$.
The same method is applied for negative particles, i.e., $\overline{p}$/$\pi^-$
below 2\,GeV/$c$ and $\overline{p}/\pi^0$ above 1\,GeV/$c$. Lines show the
$p_T$-independent systematic error bands, and the error bars show the
quadrature sum of statistical and $p_T$-dependent systematic errors. Both
$p/\pi$ and $\overline{p}/\pi$ ratios show a clear centrality dependence. In central collisions,
they reach to unity at 2$\sim$3\,GeV/$c$, while in peripheral collisions the ratios saturate at a level $\approx 0.3 - 0.4$.  

\begin{figure}[htbp]
\begin{center}
\vspace{-7mm}
\epsfig{file=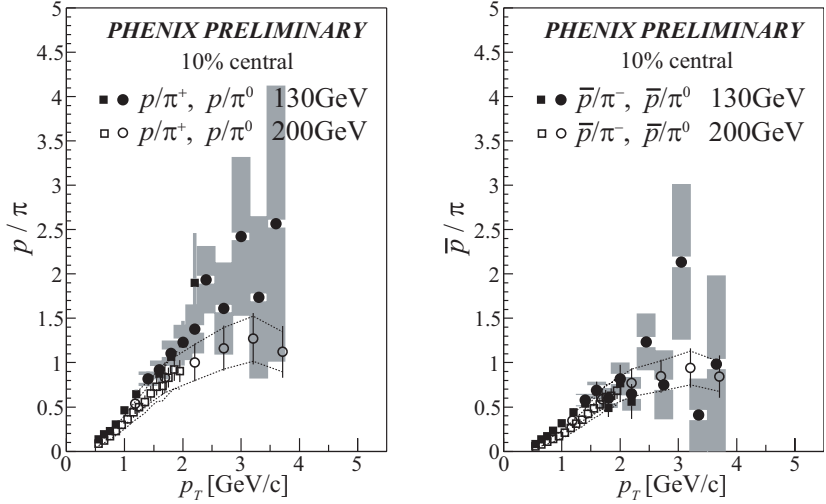,width=11cm}
\vspace{-10mm}
\caption{Comparison of $p/\pi$ and $\overline{p}/\pi$ ratios with the 130 GeV result for 0-10\,\% central collisions. The definition of the errors are same as Fig.~\ref{fig:P_Pi}. \label{fig:compare130_200}}
\vspace{-8mm}
\end{center}
\end{figure}

Figure~\ref{fig:compare130_200} shows the comparison of  0-10\,\% central $p/\pi$ and
$\overline{p}/\pi$ ratios with the 130 GeV result.
The trend observed at 130 GeV is also evident in the higher statistics 200 GeV data, although 
the $p/\pi$ ratio at 200 GeV is systematically lower. The present level of statistical and 
systematic errors does not provide a definite conclusion on whether or not the relative 
proton contribution to the hadron spectra in central Au-Au collisions changes with beam energy.
This observation, however, is consistent with the independently measured inclusive charged hadron spectra~\cite{jia}.   

\section{Conclusion}
The PHENIX detector at RHIC measured high $p_T$ identified hadron ratios in
$\sqrt{s_{NN}}$ = 200\,GeV Au+Au collisions. Within the current systematic and
statistical errors, $\overline{p}/p$ ratios that are measured up to 3.8\,GeV/c
are almost independent of both $p_T$ and centrality. The $p/\pi$ and $\overline{p}/\pi$ ratios, 
on the other hand, show strong centrality dependence. In central collisions, the (anti)proton yields 
are comparable to the pion yields at $p_{T}$ 2$\sim$4\,GeV/$c$, while for peripheral collisions 
$p/\pi$ and $\overline{p}/\pi$ saturate at $\approx 0.3 - 0.4$ .

\end{document}